# Design of optimal illumination patterns in single-pixel imaging using image dictionaries

Jun Feng,[1] Shuming Jiao,[1,2] Yang Gao,[1] Ting Lei,[1] Xiaocong Yuan[1,*]

[1]Nanophotonics Research Center, Institute of Microscale Optoelectronics, Shenzhen University, Shenzhen, Guangdong, 518060, China
[2]albertjiaoee@126.com
*Corresponding author: xcyuan@szu.edu.cn

**Single-pixel imaging (SPI) has a major drawback that many sequential illuminations are required for capturing one single image with long acquisition time. Basis illumination patterns such as Fourier patterns and Hadamard patterns can achieve much better imaging efficiency than random patterns. But the performance is still sub-optimal since the basis patterns are fixed and non-adaptive for varying object images. This Letter proposes a novel scheme for designing and optimizing the illumination patterns adaptively from an image dictionary by extracting the common image features using principal component analysis (PCA). Simulation and experimental results reveal that our proposed scheme outperforms conventional Fourier SPI in terms of imaging efficiency.**

http://dx.doi.org/10.1364/OL.99.099999

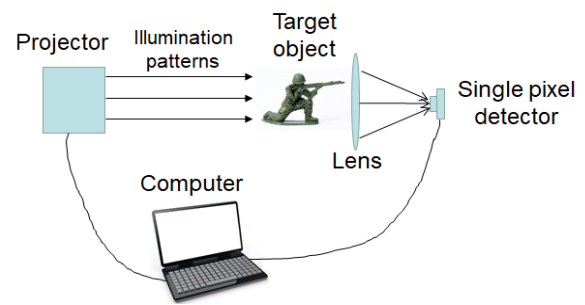

**Fig. 1.** Optical setup for a single-pixel imaging system.

Single-pixel imaging (SPI) [1] is a novel computational imaging technique for recording an object image using a single-pixel detector without spatial resolution. In conventional imaging, a two-dimensional object image is recorded at once by a pixelated image sensor array. In SPI, varying illumination patterns are projected onto the object sequentially and a single-pixel intensity sequence is recorded correspondingly. Mathematically, each single-pixel intensity value is the inner product between the object image and each illumination pattern. Finally the two-dimensional object image can be computationally reconstructed when both the recorded intensity sequence and the illumination patterns are known. A typical optical setup of SPI is shown in Fig. 1.

SPI can exhibit substantial advantages over conventional imaging under some circumstances, e. g. when the cost of pixelated image sensor is high for certain spectral bands, when no direct line of sight is available for the object scene or when the light condition is very weak. In previous works, SPI has been extensively investigated for various applications such as terahertz imaging [2], three dimensional (3-D) imaging [3,4], microscopy [5], scattering imaging [6], optical security [7,8], and gas leak monitoring [9].

Despite the success, some challenges remain to be addressed for SPI systems. One inherent drawback of a SPI system is its low imaging efficiency and long image acquisition time. A reconstructed object image with reasonable quality can only be obtained when an adequate number of illuminations are sequentially projected and the corresponding single-pixel intensities are recorded by the detector. It will be favorable if a high-quality object image can be reconstructed using only a small number of illuminations. In previous works, the imaging efficiency of SPI is mostly improved from two aspects: optimized reconstruction algorithms and optimized illumination patterns. Compressive sensing [10-12] is one common reconstruction algorithm for random illumination patterns to achieve high-quality reconstruction at a low sampling ratio in SPI .

On the other hand, the imaging efficiency can be improved by designing proper basis illumination patterns such as Fourier transform sinusoidal patterns [13,14], Hadamard transform patterns [15], cosine transform patterns [16], and wavelet transform [17] patterns, instead of random intensity patterns. For a natural smooth image, the low frequency components in the transformed domain (e.g. Fourier transform, Hadamard transform, and cosine transform) account for most information. The object image can be approximately reconstructed from only the small number of single-pixel intensities corresponding to these low frequency components. Therefore a SPI with basis patterns can achieve the same imaging quality with a significantly reduced number of illuminations. In addition, the image reconstruction can usually be realized with a fast inverse orthogonal transform for basis-pattern illumination, instead of iterative optimization in compressive sensing.

In many cases, a SPI system is usually applied to a specific scenario. The target object images may belong to a specific category and have some common features. The basis patterns in the previous schemes [13-17] are fixed for varying categories of target object images and hence the imaging efficiency is still sub-optimal. In this paper, optimized illumination patterns are designed for each different category of object images adaptively by extracting the common image features from a pre-given image dictionary with the principal component analysis (PCA) algorithm. It shall be noticed that, the concept of image dictionary and the concept of PCA (or singular value decomposition, SVD) have already been attempted in previous SPI works [12, 18]. But these works [12, 18] focus on the optimal reconstruction algorithms when the illumination patterns are random intensity patterns in SPI. The illumination patterns in SPI can be optimized as one layer in the deep learning network but only modified Hadamard binary patterns can be obtained [19]. The illumination patterns can be optimized from sample images for object classification but it is not for real imaging [20]. In the previous work [21], the illumination patterns in SPI are designed from a dictionary under a compressive sensing framework, which requires some computational cost of iterative optimization for image reconstruction [22]. In comparison, a non-iterative inverse orthogonal-basis transform can be directly employed for image reconstruction in this work. Overall, an optimized basis-illumination pattern design scheme based on image dictionaries for SPI with PCA is proposed for the first time in this work.

It is assumed that the spatial resolution of each illumination pattern, or the total number of pixels, is $N = X \times Y$. In fact, this resolution is identical to the object image resolution in the imaging model. In our proposed scheme, a number of exemplar object images (training images) need to be collected in advance to constitute an image dictionary for a specific imaging scenario. Then the optimal illumination patterns are designed based on the common features extracted from the image dictionaries and can be applied to any target object image similar to the exemplar images in the dictionary. It is assumed that totally $M$ training images are used in the dictionary and each image is represented by a row vector of length $N$. A two dimensional matrix $D$ with size $M \times N$ can be employed to represent all the training images. A normalized dictionary matrix $A$ can be obtained after the average value of each column vector in $D$ is subtracted from each element in the corresponding column. Then, the covariance matrix $A^T A$ with a size of $N \times N$ is calculated. A PCA [23-26], or SVD, can be performed on $A^T A$ to decompose the matrix into eigenvalues and eigenvectors. It is supposed that $Q$ is the eigenvector matrix of $A^T A$ with size $N \times N$ (each row is an eigenvector). The eigenvectors (rows) in $Q$ are arranged in the descending order of corresponding eigenvalues.

From a statistical point of view, PCA is a transform that converts a set of observations of $M$ correlated variables into a weighted combination of $N$ orthogonal uncorrelated basis variables (principal components). Here in our case, each variable is equivalent to an object image and each observation is equivalent to a pixel intensity in the image. PCA can perform data reduction since the first $K$ ($1 \leq K \leq N$) principal components (eigenvectors and eigenvalues) can account for most of the variability in the original $M$ variables. In previous works, PCA has been extensively employed in data compression and dimension reduction applications [23-26].

Each row vector in $Q$ is orthogonal to each other and all the row vectors are similar to the basis patterns in Fourier transform, Hadamard transform or other transforms. For a given object image, denoted by a column vector $I$ with length $N$, the following transform can be performed given by Eq (1) and a set of weighting coefficients $W = [w_1, w_2, w_3, \Lambda\ w_N]^T$ can be obtained.

$$W = QI \qquad (1)$$

Each element in $W$ is obtained by the inner product between the column vector $I$ representing the object image and the row vector representing each principal component (or eigenvector basis pattern) $q_k (1 \leq k \leq N)$ in $Q$. This transform exactly matches with the SPI model. If $q_k$ is employed as the illumination pattern, the weighting coefficient $w_k$ can be recorded by the single-pixel detector. It shall be noticed that both the object image and illumination patterns are two-dimensional rather than a one-dimensional vector in a real SPI system and the 1D/2D conversion is a straightforward re-indexing. To achieve high imaging efficiency, only $K$ most significant principal components (instead of all $N$ ones) can be employed as the actual illumination patterns for SPI. For one target object image that does not belong to the image dictionary but has similar image characteristics to the training images, the acquired weighting coefficients for a small number of principal component patterns can already recover most image information. Finally, the target object image $I$ can be computationally reconstructed from the acquired single-pixel intensity values (weighting coefficients) $W$ and the corresponding illumination patterns $Q$ by an inverse transform, similar to Fourier transform or Hadamard transform.

The pixel intensity values in the principal components can be both positive and negative. However, illumination patterns only allow positive pixel values in a practical optical system. So the pixel values need to be normalized first. For each normalized illumination pattern, both the original pattern and the complementary pattern will be projected. For example, one pixel has a normalized intensity 0.43 and its intensity will be 1-0.43=0.57 in the complementary pattern. The final single-pixel intensity value is obtained by subtracting the single-pixel intensity value corresponding to the complementary pattern from the single-pixel intensity value corresponding to the original pattern.

In this work, four different categories of object images, including number digit images from the MNIST dataset [27], mixed fashion images (consisting of ten sub-categories) from the Fashion-MNIST dataset [28], trousers images from the Fashion-MNIST dataset [28] and boot images from the Fashion-MNIST dataset [28] are assumed to be the original object images in SPI. Some examples of images belonging to each category are shown in Fig. 2. Four different sets of illumination patterns are designed for each category of object images correspondingly by extracting the principal components with our proposed scheme described above. The first ten optimized illumination patterns for each category are shown in Fig. 3 as examples. The number of training images and testing images taken from each category is presented in Table 1.

**Table 1. Number of training and testing images for each object image category in the simulation**

| Category | Image Size | Number of training images | Number of testing images |
|---|---|---|---|
| Hand-written digit images | 28×28 | 3000 | 100 |
| Mixed fashion images | 28×28 | 1000 | 120 |
| Trousers images | 28×28 | 900 | 100 |
| Boot images | 28×28 | 980 | 120 |

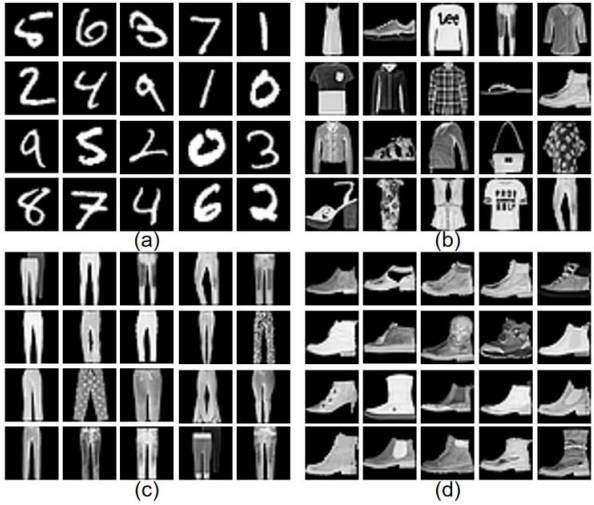

**Fig. 2.** Some examples in each category of training and testing images: (a) Number digit images from the MNIST dataset [27]; (b)Mixed fashion images from the Fashion-MNIST dataset [28]; (c)Trousers images from the Fashion-MNIST dataset [28]; (d)Boot images from the Fashion-MNIST dataset [28].

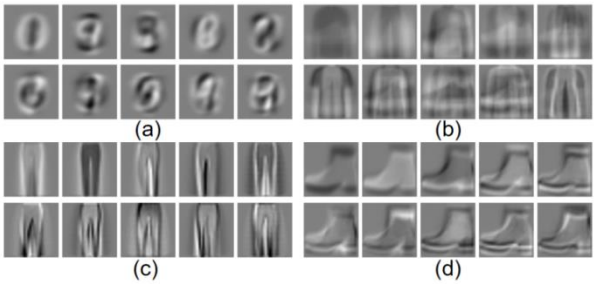

**Fig. 3.** First ten principal illumination patterns designed in our scheme for (a)number digit images, (b)mixed fashion images, (c)trousers images, (d)boots images.

In the simulation, the average quality of reconstructed testing images in the four categories with our proposed scheme is presented in Fig. 4. The results are compared with the conventional Fourier single-pixel imaging (FSPI) [13]. In FSPI, the illumination patterns are ordered from low-frequency components to high-frequency components in a zig-zag manner in the Fourier spectrum. To enhance the contrast of reconstructed images, the pixels values below a certain lower bound and above a certain upper bound are truncated.

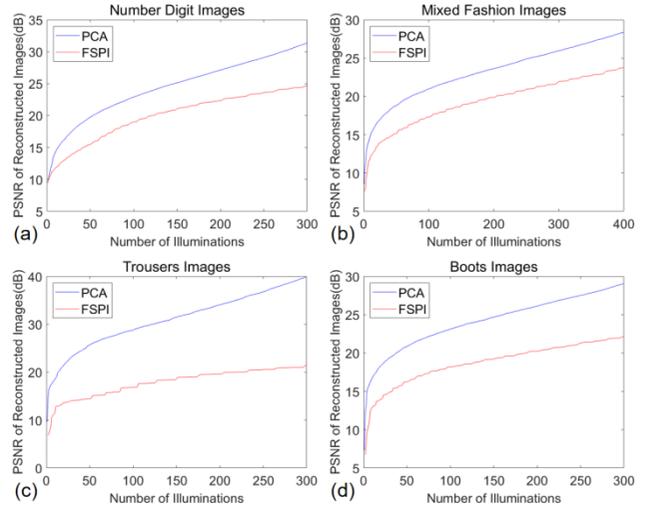

**Fig. 4.** Average image quality (PSNR) of reconstructed results in SPI with our proposed PCA scheme (blue) and conventional FSPI (red): (a)number digit images; (b)mixed fashion images; (c)trousers images; (d)boot images.

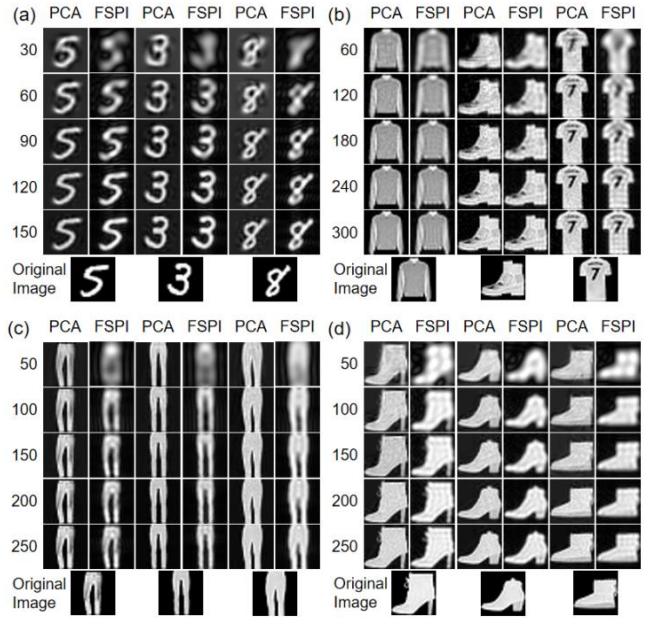

**Fig. 5.** Examples of reconstructed test images for each category of images in SPI with our proposed scheme (PCA) and conventional FSPI for varying number of illuminations (denoted by the row index): (a)number digit images; (b)mixed fashion images; (c)trousers images; (d)boot images.

It can be observed that the reconstruction results in our scheme have higher Peak-Signal-to-Noise-Ratio (PSNR) values than the ones in conventional FSPI especially at lower sampling ratios. The results in Fig. 4. indicate that our proposed scheme can yield better imaging efficiency than conventional FSPI. Some examples of reconstructed images at different sampling ratios with these two schemes are demonstrated in Fig. 5. The results in Fig. 5 visually reveal the fact that the visual quality of reconstructed object images in our proposed scheme is superior to the ones in FSPI. When the number of illuminations is small (e.g. less than 100), the reconstructed images in FSPI are usually heavily blurred but the reconstructed images already show acceptable quality in our proposed PCA scheme. The advantage of our proposed scheme is more evident for the last two categories compared with the first two categories. This is because there are more common features among the images in the last two categories than the ones in the mixed number or mixed fashion category. The performance of our proposed scheme depends on the degree of similarity between the images in the specific imaging scenario.

Optical experiments are conducted to verify the performance of our proposed scheme as well. The optical setup is shown in Fig. 6. A paper card with printed object image is illuminated by the patterns projected from a JmGO G3 projector. The single-pixel intensity values are recorded by a Thorlabs FDS1010 photodiode detector. The reconstruction results with our proposed scheme and conventional FSPI from the experimentally recorded data are shown in Fig. 7. The experimental results basically agree with the simulation results, with some extra noise contamination.

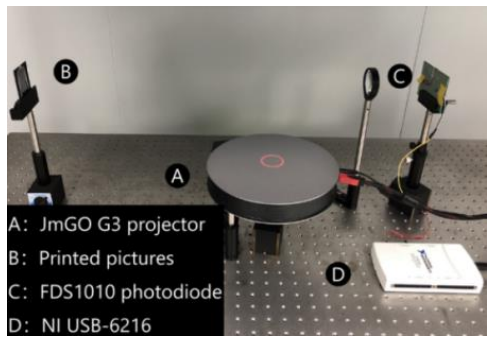

**Fig. 6.** Optical setup of our SPI experiment.

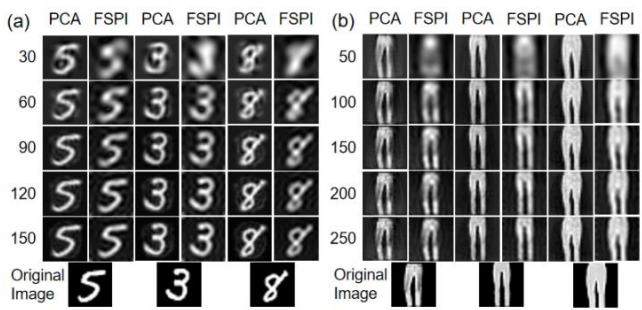

**Fig. 7.** Examples of reconstruction results for two image categories in SPI with our proposed scheme (PCA) and conventional FSPI for varying number of illuminations: (a)number digit images;    (b)trousers images.

In summary, an optimized illumination pattern design scheme in single-pixel imaging (SPI) based on image dictionaries and principal component analysis (PCA) is proposed. Compared with conventional fixed and non-adaptive basis patterns (e.g. Fourier or Hadamard patterns), the orthogonal basis patterns are adaptively optimized based on the common features of training images. Simulation and experimental results show that our proposed scheme can achieve better imaging quality than the conventional FSPI scheme under the same sampling ratio.

**Funding.** National Natural Science Foundation of China (61805145, 11774240); Leading talents of Guangdong province program (00201505); Natural Science Foundation of Guangdong Province (2016A030312010).

**Disclosures**. The authors declare no conflicts of interest.